\documentclass[12pt]{article}

\usepackage{epsfig}

{\end{list}}
\newcounter{enumct}

\begin{document}
 
\sloppy

\begin{flushright}            
DTP-99/40~~~~ \\            
April 1999~~~~ \\            
\end{flushright}
\vspace{0.6cm}  
\begin{center}
{\LARGE\bf RIDDLE OF THE SCALARS:}\\[5mm]
{\LARGE\bf WHERE IS THE $\sigma$~?}\\[10mm]
{\Large{M.R. Pennington}}\\[3mm]
{\it Centre for Particle Theory, University of Durham,}\\[1mm]
{\it Durham DH1 3LE, U.K.}\\[1mm]
{\it E-mail: m.r.pennington@durham.ac.uk}\\[10mm]

{ABSTRACT}\\[1mm]
\begin{minipage}[t]{150mm}
\baselineskip=14.5pt
The riddle of the $\sigma$ is recast in a way that tries 
to differentiate  {\it fact} from {\it fiction} as a basis for future/further discussion. By doing this, it is hoped that the
role of the $\sigma$ as dominating the ubitiquitous $\pi\pi$ interactions below 1 GeV and its relation to the QCD vacuum
can be clarified. 
\end{minipage}\\[4mm]

\end{center}
\vspace{0.3cm}

\parskip=2mm
\baselineskip=6.5mm
\section{Riddle of the $\sigma$ : what is the $\sigma$~?}

\noindent In this talk, I will start to answer some of the questions that Lucien Montanet listed in his introduction and will not attempt to survey all the known 
scalars, as this will be covered in the many other talks in this session.
I will try to differentiate clearly between those statements that
are matters of fact and those that are model-dependent and so might, at the moment, be regarded as matters of opinion. 

\noindent Let me begin with the first fact.  As is very well-known, nuclear forces are dominated by 
one pion exchange. The pion propagator is accurately described by
$1/(m_{\pi}^2-t)$, where the pion mass $m_{\pi}$ is very nearly a real number, since pions are stable as far as the strong interactions are concerned. 
The next most important contributors to nuclear forces are two pion exchange, where the pions are correlated in either an $I=1$ $P$--wave or an $I=0$ $S$--wave.  The former we know as $\rho$--exchange, the propagator for which
is described simply by ${\cal M}_{\rho}^2-t$, where now ${\cal M}_{\rho}$ is a complex number, reflecting the fact that the $\rho$ is an unstable particle.
Indeed, typically we may write ${\cal M}_{\rho}^2\,\equiv\,m_{\rho}^2\, -\, i m_{\rho} \Gamma_{\rho}$ with the mass $m_{\rho}$ and width $\Gamma_{\rho}$ real numbers.
Two pions correlated in an $S$--wave we call the $\sigma$.  However, it is an open question, whether this can be described  by
a simple Breit-Wigner propagator $1/(m_{\sigma}^2-t-im_{\sigma}\Gamma_{\sigma})$.

\begin{figure}[t]
\begin{center}
~\epsfig{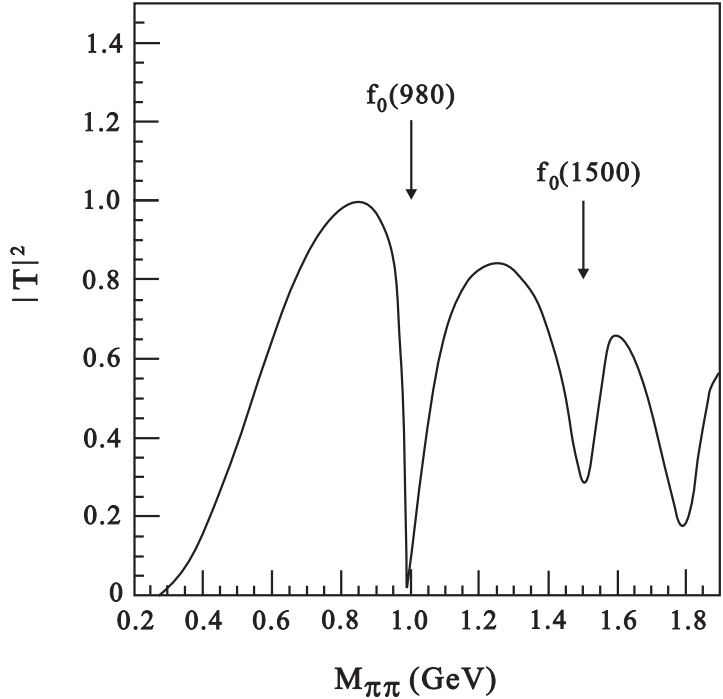}
\caption{ A sketch of the square of the modulus of the $I=0$ $\pi\pi$ $S$--wave amplitude, from Zou~$^{1)}$.} 
\end{center}
\vspace{-7mm}
\end{figure}
\noindent A hint, that the situation is not so straightforward, is given by looking in the direct channel at the $I=0$ $\pi\pi$ $S$--wave cross-section. This is sketched in Fig.~1.
One sees immediately that there are no simple Breit-Wigner-like structures.
The only narrow features are the dips that correspond to the $f_0(980)$ and the $f_0(1510)$. Otherwise we only see broad enhancements.  One might then think that there really is no sign of the $\sigma$ as a short-lived particle. Indeed, that there is no $\sigma$ has been argued by noting that this cross-section can be largely explained by $\rho$--exchange in the cross-channel. Though this is a fact, it does not immediately imply that there is no $\sigma$ in the direct channel.  Thirty years ago we learnt that Regge exchanges in the $t$ and $u$--channels not only provide an economical description of hadron scattering cross-sections above a few GeV, but that their extrapolation to low energies averages the resonance
(and background) contributions in a way specified by (finite energy sum-rule) duality. In the case of the $\pi^+\pi^-\to\pi^0\pi^0$ channel (studied recently by the BNL E852 experiment~$^{2)}$), one has just $I=1$ and $I=2$ exchanges in the cross-channel.
These Regge contributions are dominated by the $\rho$. This exchange  not only averages the
direct channel cross-section, but because there are no narrow structures near threshold, it almost equals it. If resonances are not narrow, global duality becomes local.  What we learn from this duality is that $t$--channel exchange
equates in some sense to $s$--channel resonances. Both are true. Thus,
an $s$--channel $\sigma$--resonance may well occur. (See a further comment on this in Sect.~5.)

\baselineskip=6.85mm

\noindent Why are we worried about this~? Is the $\sigma$ not just another particle in the hadron zoo~? The reason the $\sigma$ is important is because of its key role in chiral symmetry breaking~$^{3)}$.
We believe as a fact that QCD is the underlying theory of the strong interaction. The light quark sector of this theory has a chiral symmetry. The current masses of the {\it up} and {\it down} quarks are very much less than $\Lambda_{QCD}$ and to first (or zeroth approximation) are zero. It is the masses in the QCD Lagrangian that couple left and right-handed fields, so that if there are no masses, the theory has a left--right symmetry. This chiral symmetry is however not apparent at
the hadron level : pseudoscalar and scalar particles are not degenerate.
This we understand as being due to the breakdown of this chiral symmetry in the Goldstone mode~$^{4)}$, in which the scalar field acquires a non-zero vacuum expectation value, while the pseudoscalar fields remain massless. We regard pions as these Goldstone bosons, and the $\sigma$ or $f_0$ as the Higgs of the strong interaction.
It is this particle that reflects the dynamical generation of constituent masses for the {\it up} and {\it down} quarks, and so is responsible for the mass of all light flavoured hadrons. Thus the $\sigma$ or $f_0(400-1200)$ is a fundamental
feature  of the QCD vacuum. Moreover, the Goldstone nature of pions is reflected in the fact that though pions interact strongly, their interaction is weak close to threshold and so amenable to a Taylor series expansion in the low energy region --- this underlies Chiral Perturbation Theory~$^{5)}$.

\section{Where is the $\sigma$~?}

\noindent If the $\sigma$ is so fundamental, how can we tell whether it exists~?
First we recall the key aspect of a short-lived particle.  At its basic,
such a resonance gives rise to a peak in a cross-section for scattering
with the appropriate quantum numbers. Importantly, this is described in an essential way  by a Breit-Wigner amplitude, which has a pole on the nearby unphysical sheet (or sheets). It is in fact this pole that is the fundamental definition of a state in the spectrum of hadrons, regardless of 
how the state appears in experiment.  In the case of a narrow, isolated resonance, there is a close
connection between the position of the pole on the unphysical sheet and the peak we observe in experiments
at real values of the energy.  However, when a resonance is broad, and overlaps with other resonances, then this close connection is lost.  It is the position of the pole that provides the fundamental, model-independent, process-independent parameters.  While a relatively long-lived state like the $J/\psi$ appears almost the same in every channel, the $\rho$, for example, has somewhat different {\it mass} and {\it width} in different channels. This problem was recognised by the PDG long ago. 
\newpage
\baselineskip=7mm

\noindent In 1971, the $\Delta(1236)$, as it was then called,
 had been seen in many different channels. Different Breit-Wigner parameters
were noted and the PDG tables~$^{6)}$ stated: {\it We conclude that mass and width of $\Delta(1236)$ are in a state of flux; therefore we do not quote any errors in the table}. A year later~$^{7)}$, it was recognised {\it that this problem is removed if we take the mass and width to be given by the  actual pole position of the $\Delta(1236)$ in the complex energy plane.} By analytically continuing into the complex plane to the nearby pole, it was found that the pole's position was 
essentially process-independent and parameterization-independent, as $S$--matrix principles require.  Though this was known more than 25 years ago, it has often been forgotten. For the $\rho$, the 1998 PDG tables~$^{8)}$ quote a mass and width determined as Breit-Wigner parameters on the real axis. These are displayed in the complex energy plane as $E\,=\,M\,-\,i\Gamma/2$. By expanding the relevant
region, we can plot these real axis parameters as shown in Fig.~2. The points are scattered about.
However, if one now analytically continues into the complex plane, one finds that these correspond~$^{9)}$ to the pole mass and width plotted as 
$\otimes$. One sees that these
 concentrate together 10--12 MeV lower than the real axis parameters.  It is these pole parameters that are the
closest present data gets to the true parameters of the $\rho$--resonance.

\begin{figure}[p]
\begin{center}
~\epsfig{file=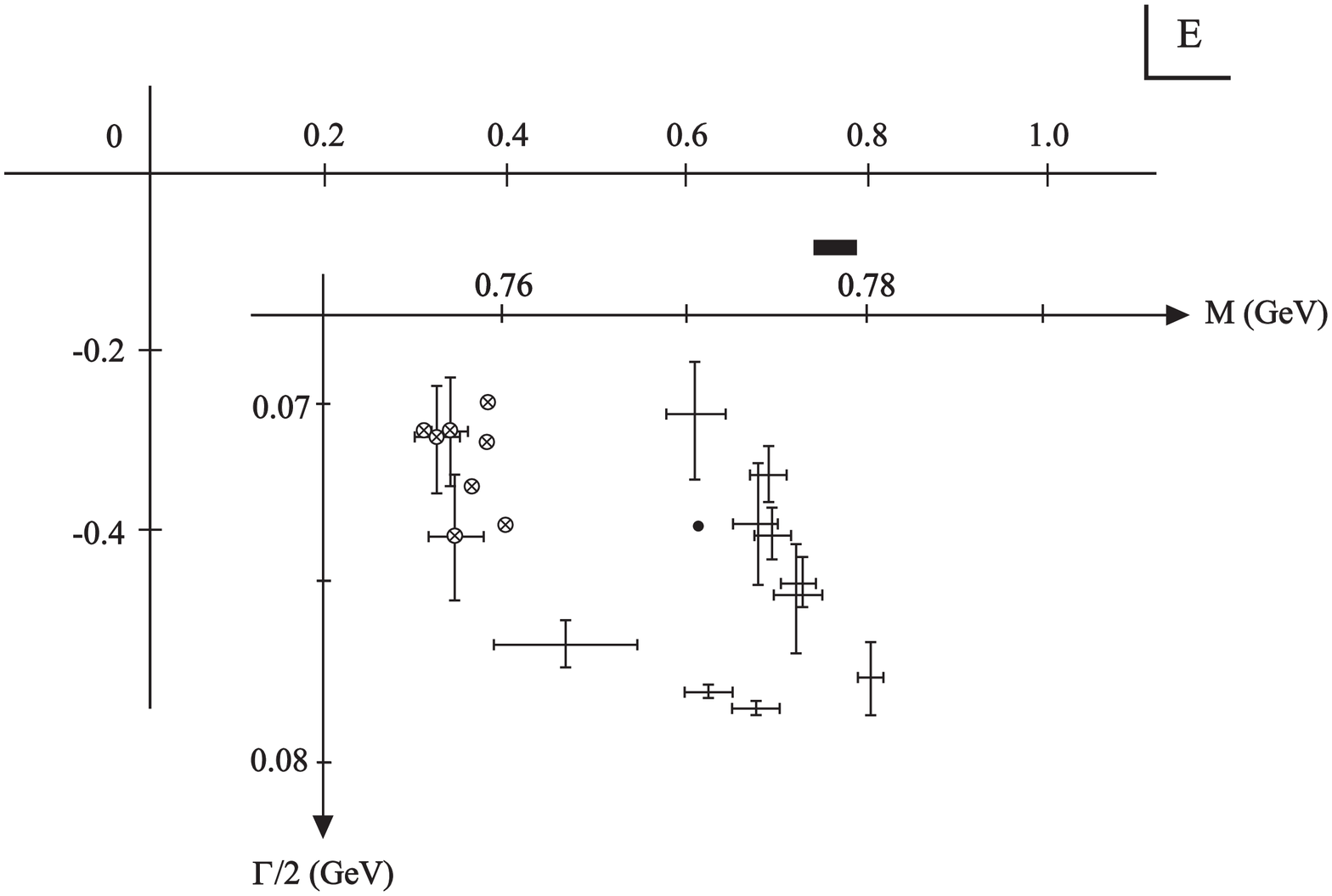,width=13cm}
\caption{ The complex energy plane is displayed. The section with data-points is an enlargement of the shaded rectangle in the bigger plot. The Breit-Wigner
mass and width parameters and their corresponding pole mass and width parameters are plotted.  The pole positions are differentiated by being shown as $\otimes$. The references can be found in the paper by Benayoun {\it et al.}~$^{9)}$} \end{center}
\begin{center}
~\epsfig{file=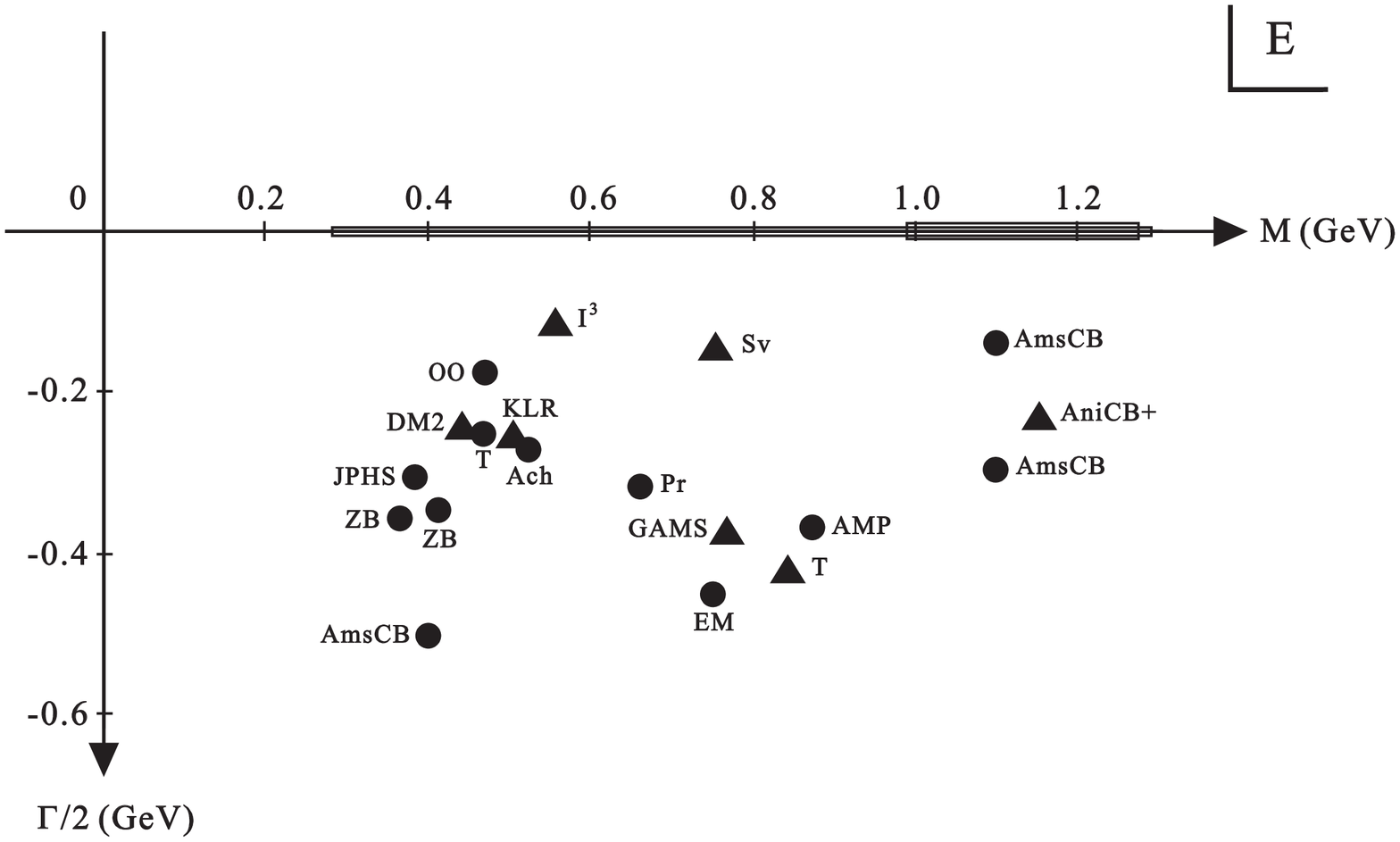,angle=0,width=13cm}
\caption{ The complex energy plane is displayed. The
mass and width of the $\sigma$ determined in each of the analyses listed in the PDG tables~$^{8)}$ is shown. The letters next to the symbols indicate the authors responsible for the analysis. See the PDG listings~$^{8)}$ for the complete list of references.} 
\end{center}
\vspace{-5mm}
\end{figure}

\noindent Now let us turn to the $\sigma$. In Fig.~3 are shown the
mass and width from the determinations given in the  1998 PDG Tables~$^{8)}$. The labels
correspond to the initials of the authors given there. Only the circles are from attempts to determine the pole positions; the triangles are Breit-Wigner-like modellings. One sees that where
the $\sigma$ is is quite unclear. Its mass is anything from 400 MeV to
1200 MeV and its width from 200 MeV to 1 GeV. The reason for this is not hard to see. The parameters  only become model-independent when close
to the pole, as we illustrate below. In a very hand-waving sense, the accuracy with which one can continue into the
complex plane is governed by the range and precision with which one knows the amplitude
along the real axis. Even for the $I=1$ $P$--wave, where the precision is good and the pole not far (some 70 MeV) from the real axis, there is a shift of 10--12 MeV. For the $I=0$ $S$--wave, any pole may be 200--500 MeV away and the precision, with which this component of $\pi\pi$ scattering, is known is not very good, Consequently, for the $\sigma$ (the $f_0(400-1200)$),
 any pole is so far in the complex plane that a continuation is quite unreliable without the aid of detailed modelling of the continuation.  Indeed, we need to differentiate strongly between the form of the amplitude on the real axis and far away near the pole. Let us do that with a simple illustration.

\newpage

\noindent Consider some scattering amplitude, ${\cal T}(s)$, for the process $1\to 2$. In the neighbourhood of the pole in the complex energy plane, where $s=E^2$, we can write
$${\cal T} (s)\,=\, \frac{g_1^R\,g_2^R}{m_R^2\,-\,s\,-\,im_R \Gamma_R}\,+\,B(s)\; , \eqno(1)$$
where the residue factors $g_1^R$ and $g_2^R$ give the coupling of the resonance to the initiating formation or production channel and to the decay channel, respectively.  Just as the mass, ${\cal M} \equiv m_R -i \Gamma_R/2$ is complex, so these couplings will, in general, also be complex. It is the pole position defined here, and the residue factors, that will be model and process-independent. Now we, of course, observe scattering only for real values of the energy. There we represent the amplitude by
$${\cal T}(s)\,=\,\frac{g_1(s) \, g_2(s)}{m^2(s)\,-\,s\,-\,i m(s) \Gamma(s)}\,+\,b(s)\; . \eqno(2)$$ 
This corresponds to  a generalised Breit-Wigner representation, in which not only, the ``width'' will be a function of $s$, but the ``mass'' too.
The Breit-Wigner mass and width are then just
\vspace{-2mm}
$$M_{BW}\,=\,m\left(s=M_{BW}^2\right)\; ,\qquad \Gamma_{BW}\,=\,\Gamma\left(s=M_{BW}^2\right)\; . \eqno(3)$$
\vspace{-7mm}

\noindent Importantly, the parameters $m(s)$, $\Gamma(s)$ and the $g_i(s)$ will not only be process-dependent, but also depend on the way the {\it background} $b(s)$ is parametrized.
However, when a pole is very close to the real axis, as in the case of the $J/\psi$,
there is essentially no difference between the pole and real axis parameters.
This is, of course, not the case for poles that are further away, even for the relatively nearby $\rho$. The parameters of Eq. (2) are connected to those of Eq.~(1) by an analytic continuation.
The functions must have the correct cut-structure to do this in a meaningful way.
For the $f_0(400-1200)$ the connection is wild and unstable, without a detailed modelling of this continuation. An example of such modelling will be given later.

\noindent It is important to realise that the unitarity of the $S$--matrix means that the pole-positions, given by Eq.~(1), transmit universally from one process to another, independently of $B(s)$. This does not hold for the real axis parameters
of Eq.~(2). Indeed, the parameters of the Breit-Wigner and background, $b(s)$,
are correlated. Thus, for instance in elastic scattering, unitarity
requires that it is the sum of the phases of the Breit-Wigner and background component that transmits universally from one process to another and not the
Breit-Wigner component separately from the background.  This fact is most
 important and one often forgotten in determining resonance parameters from Eq.~(2) and not Eq.~(1).  This is beautifully illustrated by the fits of the Ishidas and their collaborators~$^{10)}$.
\begin{figure}[h]
\begin{center}
~\epsfig{file=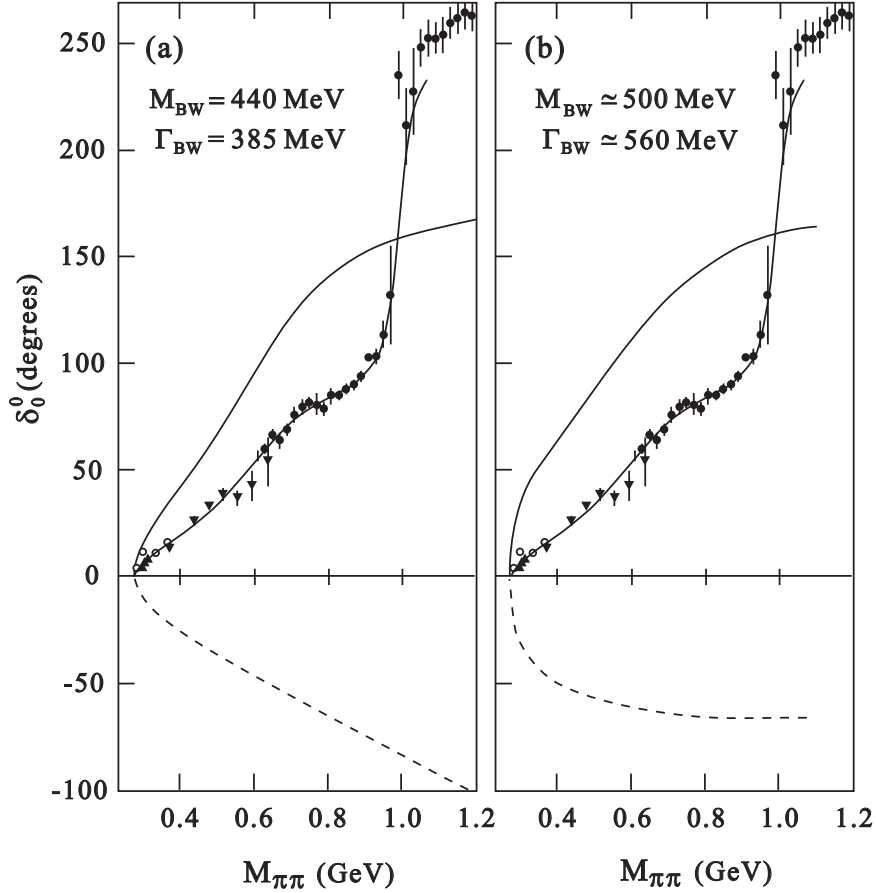,width=11.5cm}
\caption{ Two fits to the $\pi\pi$ phase-shift of Ochs~$^{11,12)}$ in terms of Breit-Wigner phase $\delta_{BW}$ (solid line) and a background component $\delta_{bkgd}$ (dashed line). Their sum is the line through the data. Example (a) is from Ishida {\it et al.}~$^{10)}$. (b) is one of an infinity of variations showing
how sensitive the Breit-Wigner mass and width, $M_{BW}$, $\Gamma_{BW}$, are to the choice of $\delta_{bkgd}$.} 
\end{center}
\vspace{-2mm}
\end{figure}
\baselineskip=6.7mm

\noindent Recognising that Watson's theorem requires that the total phase, $\delta$,
must equal the sum of the Breit-Wigner phase, $\delta_{BW}$, and
background, $\delta_{bkgd}$, they choose the background phase for isoscalar scalar $\pi\pi$ scattering to have a particular momentum dependence, shown in Fig.~4a. They then deduce the Breit-Wigner component and infer that
the parameters of Eqs.~(2,3) give
$$M_{BW}\,=\,440\,{\rm MeV}\; ,\qquad \Gamma_{BW}\,=\,385\,{\rm MeV}\; ,$$
from the fit of Fig.~4a, to the standard Ochs-Wagner phase-shifts~$^{11)}$ from the classic CERN-Munich experiment~$^{12)}$.  Since it is only the total phase, $\delta$, that matters,
one can equally choose some different background, Fig.~4b, and then deduce that
\vspace{-2mm}
$$M_{BW}\,\simeq\,500\,{\rm MeV}\; ,\qquad \Gamma_{BW}\,\simeq\,560\,{\rm MeV}\; ,
$$
\noindent for the Breit-Wigner-like component.
Of course, any other choice of background is just as good. This shows that from the real axis, one can obtain more or less any set of Breit-Wigner parameters one likes for the $\sigma$ and yet describe exactly the same experimental data. Indeed, from the analysis by Kaminski {\it et al.}~$^{13)}$ of the polarized scattering results,
described here by Rybicki~$^{14)}$, one sees that the uncertainties on the starting phase-shifts may be presently far greater than those indicated in Fig.~4.
This just makes the matter worse. The pole parameters are the only meaningful ones, but determining these directly from data on $\pi\pi$ scattering
lacks any precision.

\noindent Because of the ubiquity of $\pi\pi$ final states in almost any hadronic process,
it is useful (if not crucial) to include data from other initiating channels too.
What unifies all these is unitarity. Consider $I=0$ $J=0$ interactions for definiteness. Let ${\cal T}_{ij}$ be the amplitude for initial state $i$ to go to
final state $j$, then the conservation of probability requires that
$${\rm Im}\, {\cal T}_{ij}(s)\,=\,\sum_{n}\,\rho_n
\,{\cal T}_{in}^*(s)\,{\cal T}_{nj}(s)\quad ,\eqno(4) 
$$
\noindent
where the sum is over all channels $n$ physically accessible at the energy $\sqrt{s}$ and $\rho_n$ is the appropriate phase-space for channel $n$.  Most importantly, any channel with the same final state $j$, for instance $\pi\pi$, but initiated by a non-hadronic process, e.g.~$\gamma\gamma\to\pi\pi$, has an amplitude ${\cal F}_j$  closely related to the hadronic scattering amplitudes, ${\cal T}_{ij}$, again by the conservation of probability.  Unitarity then requires that
$${\rm Im}\, {\cal F}_{j}(s)\,=\,\sum_{n}\,\rho_n
\,{\cal F}_{n}^*(s)\,{\cal T}_{nj}(s)\quad . \eqno(5)
$$
\vspace{-5mm}

\noindent In the elastic region, where $i=j=n$, this relation becomes the well-known final state interaction theorem due to Watson~$^{15)}$, that requires the phase of ${\cal F}_i$ to be the same as the phase of the hadronic amplitude ${\cal T}_{ii}$.
It is the elastic phase-shift that transmits universally from one process to another. Unitarity knows of no separation into Breit-Wigner and background components, only the sum transmits. It is the nature of final state interactions that
when, for instance, a pion pair is produced in $\gamma\gamma$ or in $e^+e^-$
collisions, they continue to interact
independently of the way they have been produced --- only quantum numbers matter.

\begin{figure}[t]
\begin{center}
~\epsfig{file=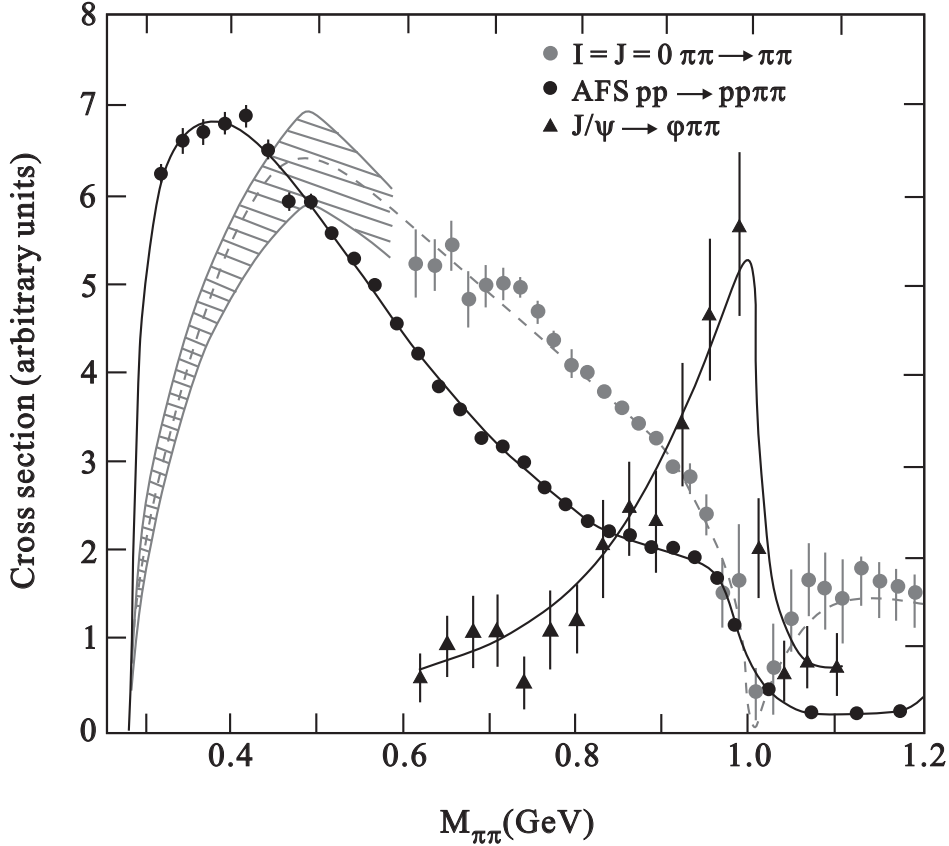,width=12.5cm}
\vspace{-2mm}
\caption{ Cross-sections for $I=0$ $S$--wave $\pi\pi\to\pi\pi$ scattering from CERN-Munich~$^{11,12)}$ and $pp\to pp \pi\pi$ from the AFS experiment at the ISR~$^{18)}$, together with the
$J/\psi\to\phi\pi\pi$ decay distribution from Mark III~$^{19)}$.  The hatched band is based on an extrapolation using the Roy equations, from Ref.~17.} 
\end{center}
\vspace{-7mm}
\end{figure}

\noindent In the multi-channel case, the solution to Eqs.~(4,5) is easily deduced~$^{16)}$ to be
$$ {\cal F}_j(s)\,=\,\sum_i\,\alpha_i(s)\,{\cal T}_{ij}(s)\quad ,\eqno(6)
$$
\noindent where the functions $\alpha_i(s)$ must be real. These determine the relative strengths of the coupling of the non-hadronic production channel to that for hadronic scattering, and are referred to as coupling functions. Eq.~(6) is an exact statement
of the content of unitarity and its universality. It ensures that any resonance in one channel couples universally to all processes that access the same quantum numbers.  Of course, it does not mean that all processes are alike!

\noindent We may treat central production of pion pairs as a quasi-non-hadronic reaction, at least in certain kinematic regimes, like high energies and very small
momentum transfers and big rapidity separations. Then the final state protons
do not interact directly with the centrally produced mesons. Similarly, the decay $J/\psi\to\phi(\pi\pi)$ is not expected to have any sizeable strong interaction between the $\phi$ and the pions. Consequently, their amplitudes for $I=0$ $S$--wave $\pi\pi$ production satisfy Eq.~(6). In Fig.~5 are shown the cross-sections for $\pi\pi\to\pi\pi$ scattering~$^{11,17)}$, $pp\to pp\pi\pi$~$^{18)}$ and the $J/\psi\to\phi\pi\pi$ decay distribution~$^{19)}$.  The difference between these is reflected in differences
in the coupling functions $\alpha_i(s)$. We see that apart from an Adler zero,
near threshold, that suppresses $\pi\pi$ elastic scattering at low energies,
central production has a very similarly shaped cross-section. In particular,
the $f_0(980)$ produces a drop or shoulder in each of them.  This is consistent with the notion that Pomerons, that supposedly control this central production process, couple to configurations of  {\it up} and {\it down} quarks, in a similar way to
$\pi\pi$ scattering. In contrast, in  the $J/\psi$ decay, the final state $\phi$ picks out hidden strangeness and so the $f_0(980)$ appears as a peak, reflecting its strong coupling to $K{\overline K}$. It is the coupling functions that shape the characteristics of these processes with the general unitarity relation of Eq.~(6) as the underlying principle.

\noindent Inexplicably, the authors of Refs.~20 have taken this universality
as implying that the coupling functions are {\it constants}. Then all the three
processes in Fig.~5 would look alike, which of course, they do not ---
for very good reason. It is the difference in the coupling functions that reveals the nature of any resonance that couples to these channels. The fact that
the $f_0(980)$ appears as a peak in $J/\psi\to\phi(\pi\pi)$, in $D_s\to \pi(\pi\pi)$
and $\phi\to\gamma(\pi\pi)$ is what teaches us~$^{16,21)}$ that the $f_0(980)$ couples strongly to $K{\overline K}\to\pi\pi$ and less to $\pi\pi\to\pi\pi$
and reflects its underlying $s{\overline s}$ or $K{\overline K}$ make-up.
The functions $\alpha_i(s)$ are not {\it meaningless} and {\it unphysical} as claimed in Refs.~20.
In their language, their parameter ${\overline \xi_f}/{\overline g_f}$  is just $\alpha(s=m_f^2)$, for instance.

\section{What is the $\sigma$~? ~$q{\overline q}$ or glueball~?}

\noindent Perhaps we can use such processes to build an understanding of the $f_0(400-1200)$ too, just as for the $f_0(980)$.  An ideal reaction in this regard is
the two photon process. As photons couple to the charged constituents of hadrons, their two photon width measures the square of their average charge squared.
We have data on both the $\pi^+\pi^-$ and $\pi^0\pi^0$ final states from Mark II~$^{22)}$, CELLO~$^{23)}$ and Crystal Ball~$^{24)}$. The underlying physics of the cross-sections
shown in Fig.~6 is reviewed in detail in Refs.~25,26.
\begin{figure}[t]
\begin{center}
~\epsfig{file=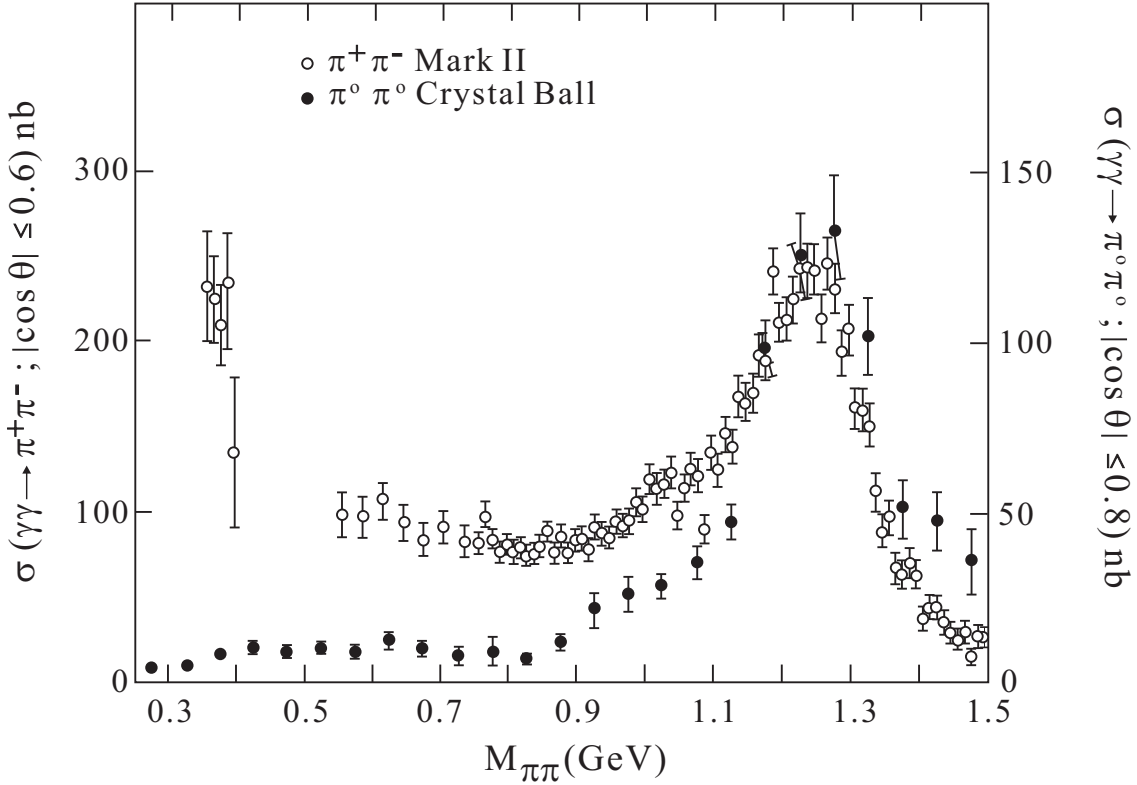,width=15.cm}
\caption{ Integrated $\gamma\gamma$ cross-section for the production of pions as a function of $\pi\pi$ mass~: charged pion data from Mark II~$^{22)}$, neutral results from Crystal Ball~$^{24)}$. } 
\end{center}
\vspace{-7mm}
\end{figure}
\noindent  Suffice it to say that the $f_2(1270)$ is most evident, but where is the $\sigma$~? Now it is often argued in the literature~$^{27)}$ that since the charged cross-section at low energies is dominated by the one-pion exchange Born term, the neutral one provides a ready measure
of the $\sigma$'s contribution. Looking at Fig.~6, this must be very small below
900 MeV. Consequently, the $\sigma$ must have a very small $\gamma\gamma$ width
and so have little of charged constituents --- perhaps it's a glueball~?
This is to misunderstand the nature of final state interactions.
These affect the charged cross-section much more dramatically than the neutral
one and this must be explained within the same modelling.
How to handle this is fully described in Refs.~25,26, so let us just deal with an essential point here.

\baselineskip=6.8mm

\noindent Imagine constructing the two photon amplitude from Feynman diagrams
and simplistically assume the contributions are just the Born term, or rather its $S$--wave component,  we call $\sqrt{2/3}\ B_S$, and the $\sigma$ contribution, $\Sigma$, incorporating the direct $\gamma\gamma$ couplings of the $\sigma$. As there is no Born contribution to the $\pi^0\pi^0$ cross-section,
it is assumed to be given wholly by this $\sigma$ component. Thus from the
measured $\gamma\gamma\to\pi^0\pi^0$ cross-section, Fig.~6, we know $\mid \Sigma \mid$.
\begin{figure}[h]
\begin{center}
~\epsfig{file=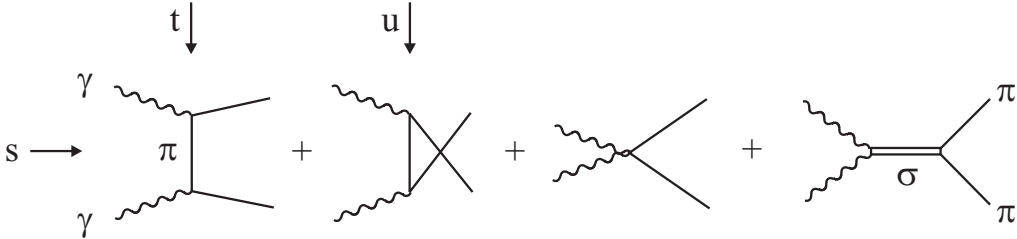,width=13.5cm}
\vspace{3mm}
\caption{ Feynman diagram modelling of $\gamma\gamma\to\pi\pi$ : the Born term plus direct channel $\sigma$ formation.}
\vspace{-7mm} 
\end{center}
\end{figure}
Similarly, by taking the measured $\gamma\gamma\to\pi^+\pi^-$ cross-section (of course, taking into account the limited angular range of such data, Fig.~6), and subtracting
the contribution from $L \ge 2$ partial waves given by the Born amplitude, one obtains the modulus of the \lq\lq charged'' $S$--wave. At 600 MeV, for instance, we have $\mid \Sigma\mid  = 0.35$ and $\mid B_S + \Sigma \mid = 0.16$, where the amplitudes are conveniently normalized (cf. Ref.~25) so that the $B_S =\sqrt{3/2}$ at threshold.\footnote{This  normalization has been chosen to avoid square roots in the labelling of Fig.~8.} These constraints are displayed in Fig.~8a. Their intersection fixes  the vector $\Sigma = \Sigma_1$.
\begin{figure}[t]
\vspace{-5mm}
\begin{center}
~\epsfig{file=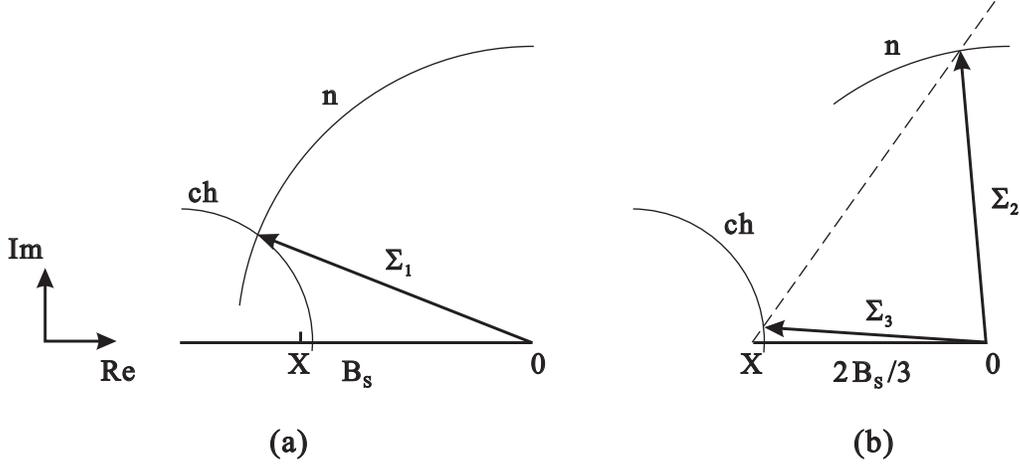,width=13.5cm}
\caption{ $B_S$ and $\Sigma$, the Born and $\sigma$ components, of the $\gamma\gamma\to\pi\pi$ $S$--wave amplitude at 600 MeV, as described in the text. The arcs of circles, labelled {\it ch, n}, are fixed by the
charged and neutral pion cross-sections, respectively, Fig.~6. They are the same circles in (a) and (b). Each $\Sigma$ vector is measured from the origin $0$.The corresponding $I=0$ $S$--wave vector runs from the point $X$ to the end of the vector $\Sigma_i$. The experimental errors on the {\it ch, n} circles have not been included for the sake of clarity: their addition does not alter the conclusions.} 
\end{center}
\vspace{-5mm}
\end{figure}
\noindent  However, final state interactions are specified by Watson's theorem, which requires  that the phase of the $\gamma\gamma\to\pi\pi$ $S$--wave amplitude
must be the same as that for the corresponding $\pi\pi$ partial wave.
 For the $I=0$ $\gamma\gamma\to\pi\pi$ amplitude, this means
  $$\tan \delta^0_0\,=\,\frac{{\rm Im} \Sigma}
{\frac{2}{3} B_S + {\rm Re} \Sigma} \quad ,
$$
\noindent which fixes the $I=0$ $S$--wave vector to lie along the dashed line in Fig.~8b running from the point X.
This constraint combined with the $\pi^0\pi^0$ cross-sections means $\Sigma = \Sigma_2$, whereas the $\pi^+\pi^-$ cross-section gives $\Sigma = \Sigma_3$.
Which is the right $\Sigma$--vector
 dramatically affects the size of the $I=0$ $S$--wave amplitude.
Clearly, $\Sigma_1, \Sigma_2, \Sigma_3$ should all be equal! This inconsistency is a sign of the inadequacy of such a simplistic model. Indeed, in terms of Feynman diagrams we must add to the graphs of Fig.~7
\vspace{2mm}
\begin{figure}[h]
\begin{center}
~\epsfig{file=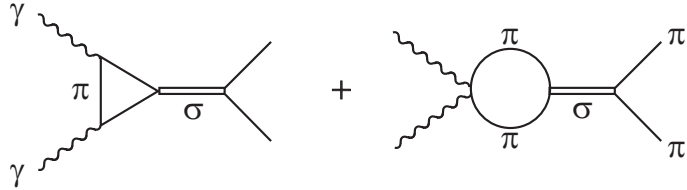,width=9cm}
\caption{ Additional contributions to the model amplitudes for $\gamma\gamma\to\pi\pi$ of Fig.~7 essential for ensuring the final state interaction theorem is satisfied.} 
\end{center}
\vspace{-8mm}
\end{figure}

\noindent as well as all the corrections to the Born term. 
Without such terms, the magnitude of the {\it direct} $\sigma$--component is meaningless. The dispersive framework sums all such terms exactly. This allows the nearest one can presently get to a model-independent separation of the individual spin components with $I=0$ and 2. This reveals a quite different
$S$--wave amplitude, see Fig.~10 for the {\it dip} solution~$^{28)}$. It is the strong interference between the contributions of Figs.~7,9 that makes the structure of the $\gamma\gamma\to\pi\pi$ $I=0$ $S$--wave quite different from that of any other process, cf. Figs.~1,10.
\begin{figure}[h]
\begin{center}
~\epsfig{file=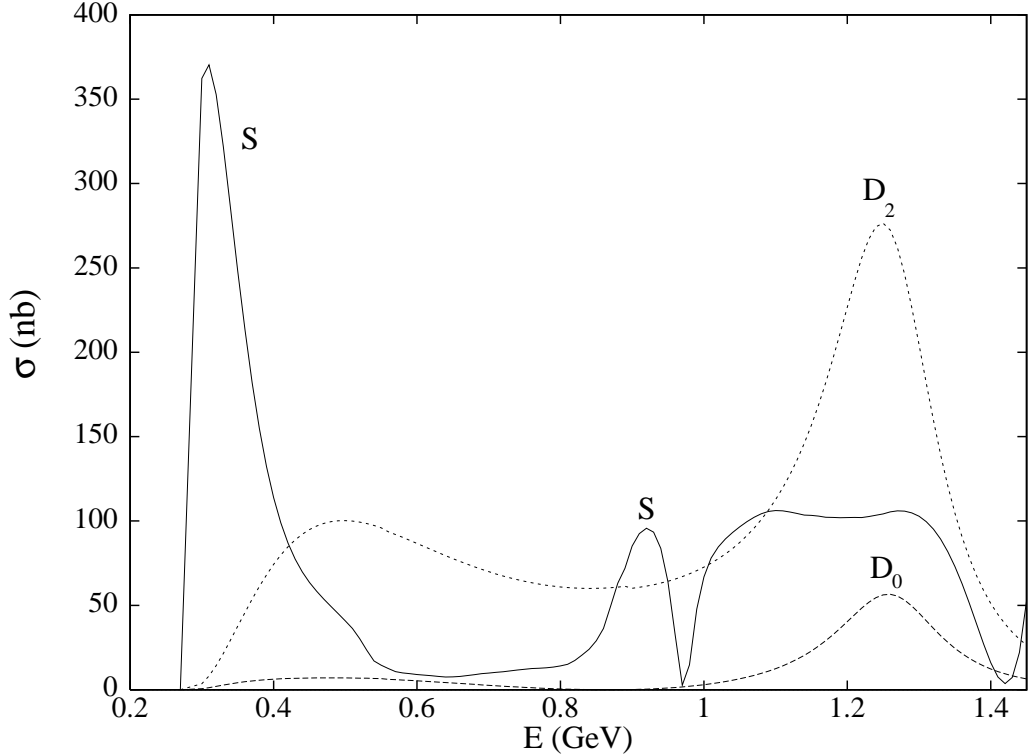,angle=-90,width=14cm}
\caption{ Integrated $I=0$ partial wave cross-sections (labelled by $J_{\lambda}$ where $J$ is the spin $J$ and $\lambda$ the helicity) for $\gamma\gamma\to\pi\pi$ from the Amplitude Analysis by  Boglione and Pennington~$^{28)}$ showing the
{\it dip} solution. See Ref.~28 for the peak solution.} 
\end{center}
\vspace{-7mm}
\end{figure}

\noindent Recalling that two photon widths are a measure of the square of the mean squared charge squared of the constituents of a hadron times the probability that these constituents annihilate, these widths tells us about the constitution of resonances. Thus,
$\Gamma(f_2(1270)\to\gamma\gamma)\,=\,(2.84 \pm 0.35)$~keV we find~$^{28)}$ is just what is expected of a $(u{\overline u} + d{\overline d})$ tensor. While
$\Gamma(f_0(980)\to\gamma\gamma)\,=\,(0.28 ^{+0.09}_{-0.13})$~keV is not only consistent with the radiative width of an $s{\overline s}$ scalar, it also
agrees with the prediction for a $K{\overline K}$ system~$^{29)}$. Indeed, the $f_0(980)$ is likely to be a mixture of both of these. For the $\sigma$ we find
$\Gamma(f_0(400-1200)\to\gamma\gamma)\,=\,(3.8 \pm 1.5)$~keV, which is quite consistent with the width expected for a 
$(u{\overline u} + d{\overline d})$ scalar, according to Li, Barnes and Close~$^{30)}$.

\noindent Another model-dependent way to test the composition of the $f_0(400-1200)$ is by the use of QCD sum-rules.
These  connect the low energy hadron world to the high energy regime of asymptotic freedom, where the predictions of QCD are calculable~$^{31)}$.
By applying sum-rule techniques to the non-strange scalar current, Cherry, Maltman and myself~$^{32)}$ have found that the $f_0(400-1200)$, as given by experiment,
cf. Fig.~1, saturates the sum-rules. This is in contradistinction to the conclusions of Elias~{\it et al.}~$^{33)}$, presented at this meeting by Steele, who find the sum-rules are not saturated, but where they describe the $\sigma$ by
a broad Breit-Wigner-like structure.

\section{Modelling the unknown --- the $\sigma$ pole}

\noindent Earlier, we have stressed the importance of the pole in determining the only truly unambiguous parameters of any short-lived state.  However, the $\sigma$ is so short-lived that continuing experimental information to the actual pole is highly unreliable without modelling. A possible way to proceed
is to approximate experiment
(at real values of the energy) by known analytic functions, one can then
readily continue to the pole. Let me illustrate this with an example. Let us consider, the scalar form-factor, $F (s)$. Though this is 
not a directly observable quantity, it has the advantage of only having a right hand cut and so its continuation is particularly straightforward.
Let us assume that, as $\mid s \mid \to \infty$, 
$$ s^{-2}\; < \; \mid F(s) \mid \; < \; s^2\quad .$$
Then both $F(s)$ and its inverse satisfy twice-subtracted dispersion relations.
\setcounter{equation}{6}
\begin{eqnarray}
F(s)&=& 1\,+\,b s\,+\,\frac{s^2}{\pi}\,\int^{\infty}_{4m_{\pi}^2}\,
ds'\;\frac{{\rm Im}\ F(s')}{s'^2\ (s' - s)}\qquad ,\\
\frac{1}{F(s)}&=& 1\,-\,b s\,+\,\frac{s^2}{\pi}\,\int^{\infty}_{4m_{\pi}^2}\,
ds'\;\frac{{\rm Im}\ 1/F(s')}{s'^2\ (s' - s)}\quad .
\end{eqnarray}

\noindent Along the real positive axis we simply model $\beta\ {\rm Im}F(s)$ by a polynomial in
$\beta^2\ =\ 1 - 4m_{\pi}^2/s$. The parameters are arranged to fulfill elastic unitarity at low energies, by fixing the phase of $F(s)$ to be the experimental $I=J=0$ $\pi\pi$ phase-shift, $\delta^0_0$ of Fig.~4.  We can then use the dispersion relations of Eqs.~(7,8) to determine
$F(s)$ everywhere in the complex energy plane $E$, where $s=E^2$. Whether one approximates the imaginary part of the form-factor, or its inverse, makes very little difference, to its continuation on the first sheet. In any event, there are no poles on this sheet.  

\begin{figure}[h]
\begin{center}
~\epsfig{file=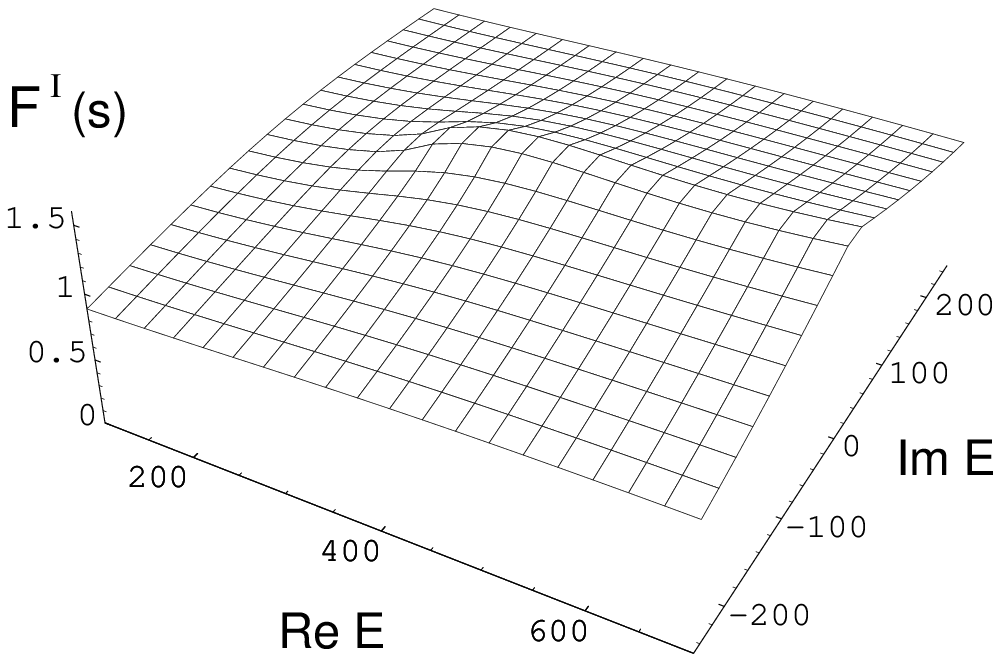,width=10.5cm}
\vspace{-4mm}
\end{center}
\begin{center}
~\epsfig{file=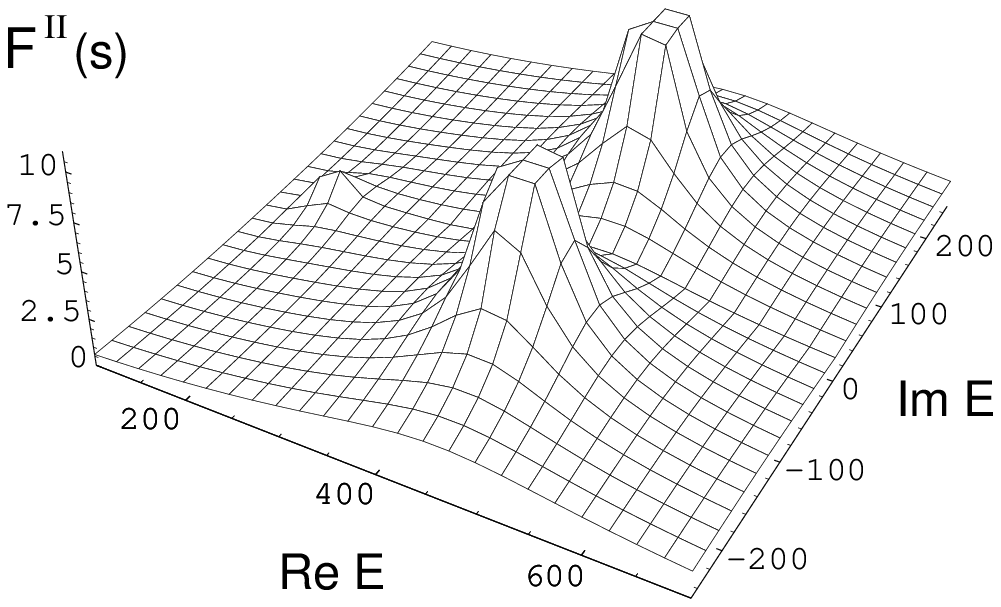,width=10.5cm}
\caption{ Modulus of the scalar form-factor in the complex energy plane $E$, with $s=E^2$,
on the first and second sheets, $F^I(s)$ and $F^{II}(s)$ respectively, computed from a twice subtracted dispersion for $1/F(s)$, Eq.~(8). $E$ is in MeV. }
\vspace{-5mm}
\end{center}
\end{figure}

\noindent We then continue to the second sheet. This is achieved by taking the other sign of the square root branch-point, i.e.
$$ \sqrt{s\,-\,4m_{\pi}^2}\;\to\; -\sqrt{s\,-\,4m_{\pi}^2}\;.$$
If one considers the continuation given by Eq.~(7), there are still no poles and $F(s)$ is smooth. However, if instead one uses Eq.~(8), poles emerge rather spectacularly, Fig.~11. This alternative approximation to the imaginary part of the inverse form-factor may be regarded as some Pad\'e approximant to the {\it exact}
imaginary part. In this simple exercise, we find the $\sigma$ pole gives, Eq.~(1),
$$M_R\,\simeq\,457\,{\rm MeV}\; ,\qquad \Gamma_R\,\simeq\,219\,{\rm MeV}\; .$$

\noindent If instead, one models the low energy form-factor by using Chiral Perturbation Theory ($\chi PT$), as has been done by by Hannah~$^{34)}$, then one finds
$$M_R\,\simeq\,463\;{\rm MeV}\; ,\qquad \Gamma_R\,\simeq\,393\;{\rm MeV}$$
at one loop $\chi PT$, and
$$M_R\,\simeq\,445\;{\rm MeV}\; ,\qquad \Gamma_R\,\simeq\,470\;{\rm MeV}$$
at two loops, making three subtractions, to emphasise still further the
low energy constraints from $\chi PT$~$^{35)}$. While the real part of the pole position is reasonably stable, the imaginary part depends rather more sensitively on the modelling of the real axis information. Dobado and Pelaez~$^{35)}$, and then Oller and Oset~$^{35)}$, discussed similar calculations first, but for the full
$\pi\pi$ scattering amplitude in $\chi PT$. These examples illustrate how precision data on $\pi\pi$ observables can determine the parameters of the $\sigma$. The precision is input by a specific  modelling that allows a continuation
giving poles on the nearby unphysical sheet.
Remember that whatever the pole parameters actually are, it is still the same
$\pi\pi$ amplitudes and phases, Figs.~1,4, that are being described.

\section{Summary --- facts}

\noindent Here we summarise the key facts discussed here:

\begin{itemize}
\item[1.] The $I=J=0$ $\pi\pi$ interaction is strong above 400 MeV, or so.
This very short-lived correlation between pion pairs is what we call the $\sigma$.

\item[2.] Such a $\sigma$ is expected to be the field whose non-zero vacuum expectation value breaks chiral symmetry. The details of this  are, however, model-dependent --- see Refs.~3,4,5,36.

\item[3.] The low mass $\pi\pi$ enhancement may be describable in terms of
$t$--channel exchanges, in particular the $\rho$, but this does not mean that the $\sigma$ does not exist as an $s$--channel pole. That there are these alternative descriptions is just hadron duality.\footnote{While an infinite number of crossed-channel exchanges are needed to generate a pole in the $s$--channel,
the $\sigma$--pole is so far from the real axis, that the absorptive part of the $I=0$ $\pi\pi$ amplitude can on the real axis be readily described by a few crossed-channel (Regge) exchanges, which is all that matters for (finite energy sum-rule) duality.}

\item[4.] It is the pole in the complex energy plane that defines the existence of a state in the spectrum of hadrons. It is only the pole position (and residues) that are model-independent.  Within models, the position of $K$--matrix poles may be imbued with significance as indicating the underlying or precursor state~$^{37,38)}$. However, these only have meaning within models and within a particular parametrization of the $K$--matrix.  In contrast, poles of the $S$--matrix are both
process and model-independent.

\item[5.] Fitting $I=J=0$ $\pi\pi$ data on the real axis in the energy plane
with Breit-Wigner forms determines $M_{BW}$ and $\Gamma_{BW}$, Eqs.~(2,3).  However, these are parametrization-dependent, process-dependent and a poor guide to the true pole position, Eq.~(1).

\item[6.] The pole position is determined by analytic continuation.
Since for the $\sigma$, this continuation is far, the mass function $m (s)$, width function $\Gamma(s)$, and couplings $g(s)$ of Eq.~(2) will all be functions of energy and not simply constant. Tornqvist has illustrated the energy dependence of such scalar propagators within a model of hadron dressing~$^{38)}$.

\item[7.] While the shape of the $\pi\pi$ spectrum is process-dependent (see Fig.~5), the phase of the corresponding amplitudes, in a given spin and isospin configuration, is process-independent below 1 GeV.
\end{itemize}

\section{Summary --- model-dependent statements}
 
\begin{itemize}

\item[1.] The link between the almost model-independent experimentally determined radiative width and the composition of the $\sigma$ does require modelling.
 Analysis of the $f_0(400-1200)$ in two photon processes indicates that it has a $(u{\overline u}\,+\,d{\overline d})$ composition.
 
\item[2.] Preliminary results from a new QCD sum-rule analysis~$^{32)}$ of the  scalar $(u{\overline u}\,+\,d{\overline d})$
current suggests that this is saturated by the $f_0(400-1200)$, just as expected from  [1] above.

\item[3.] The pole position of the $\sigma$ can be found by modelling the analytic continuation, starting from experimental (or theoretical) information
on the real axis.

\item[4.] The relation that this pole has to the underlying {\it undressed} or {\it bare} state is model-dependent.  The model of Tornqvist~$^{38)}$, for instance, provides a possible connection with the lightest underlying ideally mixed $q{\overline q}$ multiplet~$^{38,39)}$.

\end{itemize}

\noindent To go further, we need precision data on {\it understood} processes.
$\pi\pi$ final states with vacuum quantum numbers appear in a multitude of reactions. It is only by the collective analysis
 of all of these that we can hope to solve the riddle of the $\sigma$. It is a puzzle worth solving, since the nature and properties of the $\sigma$ lie at the heart of the QCD vacuum.

\vspace{1cm}
\centerline{\bf Acknowledgements}

\noindent It is a pleasure to thank the organisers, particularly
Tullio Bressani and Alessandra Filippi, for having generated
this meeting and Lucien Montanet for devoting  a day to the $\sigma$ and stimulating renewed discussion of its existence, nature and properties.
I acknowledge travel support from the EEC-TMR Programme, Contract No.
CT98-0169, EuroDA$\Phi$NE. I am especially grateful to Tony Thomas and the
{\it Special Research Centre for the Subatomic Structure of Matter} at the University of Adelaide, where this talk was prepared, for their generous hospitali

\vspace{1cm}

\centerline{\bf References}
\baselineskip=5.5mm
\parskip=1.2mm
\begin{itemize}
\parskip=0mm
\item[1.] B.S. Zou, hep-ph/9611235, Talk presented at {\it 34th Course of International School of Subnuclear Physics}, Erice, Sicily, July 1996;

V.V. Anisovich, D.V. Bugg, A.V. Sarantsev, B.S. Zou, 
Phys.Rev. {\bf D50} (1994) 972;

D.V. Bugg,  B.S. Zou, A.V. Sarantsev, Nucl. Phys. {\bf B471} (1996) 59.

\item[2.] J. Gunter {\it et al.}, hep-ex/9609010, APS Minneapolis 1996, Particles and fields vol.1 pp.~387-391.
\item[3.] M. Gell-Mann, M. Levy, Nuovo Cim. {\bf 16}, 705 (1960);

Y. Nambu, G. Jona-Lasinio, Phys. Rev. {\bf 122}, 345 (1961);

see  R. Delbourgo, M.D. Scadron, Phys. Rev. Lett. {\bf 48}, 379 (1982) in the context of QCD.
\item[4.] J. Goldstone, Nuovo Cim. {\bf 19}, 154 (1961).
\item[5.] S. Weinberg, Physica {\bf 96A}, 327 (1979);
\item[ ] J. Gasser, H. Leutwyler, Ann. Phys. (NY) {\bf 158}, 142 (1984),
Nucl. Phys. {\bf B250} 465 (1985).
\item[6.] A. Rittenberg {\it et al.}, Review of Particle Properties, Rev. Mod. Phys. {\bf 43} S1 (1971) --- see p.~S115.
\item[7.] P. S\"oding {\it et al.}, Review of Particle Properties, Phys. Lett. {\bf 39B}, no. 1 (1972) --- see p.~104.
\item[8.] C. Caso {\it et al.}, Review of Particle Physics, Eur. Phys. J. {\bf C3}, 1 (1998).
\item[9.] M. Benayoun, H.B. O'Connell, A.G. Williams, Phys.Rev. {\bf D59}, 074020 (1999).
\item[10.] S. Ishida {\it et al.}, Prog. Theor. Phys. {\bf 95}, 745 (1996),
\item[11.] W. Ochs, thesis submitted to the University of Munich (1974).
\item[12.] B. Hyams {\it et al.}, Nucl Phys. {\bf B64}, 134 (1973);
\item[ ]  G. Grayer {\it et al.}, Nucl. Phys. {\bf B75}, 189 (1974);
\item[ ] see also, D. Morgan, M.R. Pennington, {\it Second DA$\Phi$NE Physics Handbook}, ed. L. Maiani, G. Pancheri and N. Paver,  pp. 193-213 (INFN, Frascati, 1995).
\item[13.] R. Kaminski, L. Lesniak, K. Rybicki, Z. Phys. {\bf C74}, 79 (1997).
\item[14.] K. Rybicki, these proceedings.
\item[15.] K.M. Watson, Phys. Rev. {\bf 95}, 228 (1954).
\item[16.] K.L. Au, D. Morgan, M.R. Pennington, Phys. Rev. {\bf D35}, 1633 (1987).
\item[17.] M.R. Pennington, Proc. BNL Workshop on Glueballs, Hybrids and Exotic Hadrons, ed. S.U. Chung, AIP Conf. Proc. No. 185, Particles \& Fields Series 36, p.~145 (AIP, New York, 1989).
\item[18.] T. \AA kesson {\it et al.}, Nucl. Phys. {\bf B264}, 154 (1986).
\item[19.] U. Mallik, {\it Strong Interactions and Gauge Theories}, Proc. XXIth Recontre de Moriond, ed. J. Tran Thanh Van, Vol~2, p.~431 (Editions Fronti\'eres, Gif-sur-Yvette, 1986);

see also, A. Falvard {\it et al.} (DM2), Phys. Rev. {\bf D38}, 2706 (1988);

W. Lockman (Mark III), Proc {\it 3rd Int. Conf on  Hadron Spectroscopy}, Ajaccio, 1989, p.~109 (Editions Fronti\`eres, Gif-surYvette, 1989).
\item[20.] M. Ishida, S. Ishida, T. Ishida, hep-ph/9805319, Prog. Theor. Phys. {\bf 99}, 1031 (1998).
\item[ ] S. Ishida, these proceedings.
\item[21.] D. Morgan, M.R. Pennington, Phys. Rev. {\bf D48}, 1185 (1993). 
\item[22.] J. Boyer {\it et al.}, Phys. Rev. {\bf D42}, 1350 (1990).
\item[23.] H.J. Behrend {\it et al.}, Z. Phys. {\bf C56}, 381 (1992).
\item[24.] H. Marsiske {\it et al.}, Phys. Rev. {\bf D41,} 3324 (1990);
\item[ ] J.K. Bienlein, Proc. {\it IXth~Int.~Workshop on Photon-Photon
Collisions} (San Diego 1992) ed. D. Caldwell and H.P. Paar,
p.~241 (World Scientific). 

\item[25.] D. Morgan, M.R. Pennington, Phys. Lett. {\bf 192B}, 207 (1987),  Z. Phys. {\bf C37}, 431 (1988); {\bf C39}, 590 (1988); {\bf C48}, 623 (1990).
\item[26.] M.R. Pennington, {\it DA$\Phi$NE Physics Handbook}, ed. L. Maiani, G. Pancheri and N. Paver,  pp. 379-418 (INFN, Frascati, 1992);
{\it Second DA$\Phi$NE Physics Handbook}, ed. L. Maiani, G. Pancheri and N. Paver,  pp. 531-558 (INFN, Frascati, 1995).

\item[27.] S. Narison, Nucl. Phys. {\bf B509}, 312 (1998);

P. Minkowski, W. Ochs, hep-ph/9811518, EPJC (in press).

\item[28.] M. Boglione, M.R. Pennington, hep-ph/9812258, EPJC (in press).

\item[29.] T. Barnes, Phys. Lett. {\bf 165B}, 434 (1985);
 Proc. {\it IXth~Int.~Workshop on Photon-Photon
Collisions} (San Diego 1992) ed. D. Caldwell and H.P. Paar,
 p.~263  (World Scientific).

\item[30.] Z.P. Li, F.E. Close, T. Barnes,
Phys. Rev. {\bf D43} 2161 (1991).
\item[31]  M.A. Shifman, A.I. Vainshtein, V.I. Zakharov,
Nucl. Phys. {\bf B147} 385, 448 (1979).
\item[32.] S.N. Cherry, K. Maltman, M.R. Pennington, in preparation.
\item[33.] V. Elias, A.H. Fariborz, F. Shi, T.G. Steele, Nucl. Phys. {\bf A633}, 279 (1998);
\item[ ] T.G. Steele, these proceedings.
\item[34.] T. Hannah, Phys. Rev. (in press).
\item[34.] A. Dobado, J.R. Pel\'aez, Phys. Rev. {\bf D56}, 3057 (1997);

J.A. Oller and E. Oset, Nucl. Phys. {\bf A620}, 438 (1997).

\item[36.] M.R. Pennington, hep-ph/9612417, Proc. of DAPHCE Workshop, Frascati, Nov. 1996, Nucl. Phys. (Proc. Supp.) {\bf A623} 189 (1997).
\item[37.] V.V. Anisovich, A.V. Sarantsev, Phys. Lett. {\bf B382}, 429 (1996).
\item[38.] N. Tornqvist, Z. Phys. {\bf C68}, 647 (1995).
\item[39.] E. van Beveren {\it et al.}, Z. Phys. {\bf C30}, 615 (1986);
\item[ ] M. Boglione, M.R. Pennington, Phys. Rev. Lett. {\bf 79}, 1633 (1997).
\end{itemize}
\end{document}